\documentclass[doublespace,12pt,onecolumn]{ieeeconf}  
\IEEEoverridecommandlockouts           
\renewcommand{\baselinestretch}{2}

\usepackage{graphicx,comment,amsfonts,amsmath}

\newcommand{\qed}{\hfill$\Box$\\}

\def\E{{\sf E}}
\def\tE{\tilde{\sf E}}
\def\tA{\tilde{A}}
\def\P{{\sf P}}
\def\dd{\mbox{d}}
\def\me{\mbox{e}}

\def\eg{{\em e.g.},~}
\def\ie{{\em i.e.},~}

\newtheorem{proposition}{Proposition}[section]

\newtheorem{corollary}{Corollary}[section]
\newcommand{\beqa}{\begin{eqnarray*}}
\newcommand{\eeqa}{\end{eqnarray*}}
\newcommand{\be}{\begin{eqnarray}}
\newcommand{\ee}{\end{eqnarray}}

\begin{document}

\title{Relative Age of Information:\\
Maintaining Freshness 
while\\ Considering the Most Recently Generated Information}

\author{
\begin{tabular}{ccc}
George Kesidis& Takis Konstantopoulos &  Michael Zazanis\\
Penn State Univ & Univ Liverpool & AUEB\\
Univ Park, PA, USA & Liverpool, England, UK & Athens, Greece \\
gik2@psu.edu & t.konstantopoulos@liverpool.ac.uk &  zazanis@aueb.gr
\end{tabular}
}

\maketitle

\begin{abstract}
A queueing system handling a sequence
of message arrivals is considered
where each message obsoletes all previous messages.
The objective is to assess the freshness
of the latest message/information that  has been successfully
transmitted, \ie ``age of information" (AoI).
We study a variation of traditional AoI, the {\em Relative AoI},
here defined so as to account for the presence of 
newly arrived messages/information to the queue
to be transmitted.
\end{abstract}

\section{Introduction}

Certain types of communication involves a sequence of
messages where each message obsoletes all previous ones
\cite{Yates12,Eph16,Kosta17,Altman18}. 
The generated messages are sent to a transmission server.
For a simple example, a temperature sensor could periodically
transmit a reading to a remote control system. The control system
could prioritize the most recent temperature reading.
The sensor could instead be an alarm such as a motion detector,
which needs to be manually reset once tripped; any alarm message would
render stale any queued or in-transmission ``heartbeat" message 
that is periodically sent to indicate no intruder is present and that
the sensor is properly functioning.
Alternatively, the messages may be commands to a remote actuator
of a  control system.

For the transmission of such messages, an important performance
criterion is the freshness (age) of the most recently
successfully transmitted message (information), \ie
Age of Information (AoI).

Consider a stream of messages
to a transmission server 
with i.i.d. interarrival times $\sim X$ 
and i.i.d. message service-times  $\sim S$.
Again, a newly arrived message obsoletes all previous messages.
This may imply that messages are of the same or comparable length
(\ie dependent lengths) but there
may be randomness associated with service, \eg time-varying
noise or interference associated with the communication channel.

Let $D(t)$, respectively $A(t)$,
 be the largest message departure time from,
respectively message arrival time to,
the server that is $\leq t$.
Let $\tA(t)$ be the largest arrival
time among {\em successfully departed} messages
such that $\tA(t)\leq t$.
The AoI associated
with the server at time $t$ is  \cite{Yates12},
\be\label{AoI-def-old}
{\sf AoI}(t) & = & t-(\tA \circ D)(t) ~=~ t-\tA(D(t)).
\ee
Previous work on AoI has 
considered a transmission server after a lossless FIFO queue 
wherein $A=\tA$  \cite{Yates12}.
More recently, finite and lossy (e.g., bufferless)
systems and server preemption have
been considered for this definition of AoI \cite{Altman18}.
For a bufferless/queueless system under push-out considered in the following, 
$A\circ D=\tA \circ D$.
These identities will generally not hold
for systems that may block some messages upon arrival
(\ie that do not admit all messages).

Note that AoI  (\ref{AoI-def-old})
does not consider the arrival times
of new messages since the last departure. The point is that
most recently received message may be freshest possible.
With this in mind, herein we are 
interested in the non-negative difference between the AoI 
and arrival time to the transmission system of the
most recent message, \ie the {\em Relative AoI},
\be \label{DelI-def}
\Delta(t) & := & A(t)-\tA(D(t)).
\ee
That is, the question critical to performance which the
definition of $\Delta$ is trying to address is:
Has the most recently received message been transmitted?
Note that $\Delta$ is zero if this is the case.

\section{Busy periods of GI/GI/1/1-PO system}\label{sec:bg}

Consider messages sent at mean rate $\beta$
with i.i.d. interarrival times $\sim X$ such that
finite $\E X = 1/\beta>0$
and $\P(X>0)=1$.
Also, assume that the i.i.d. 
message service times $S$ 
are such that finite $\E S=1/\delta>0$ and $\P(S>0)=1$
(and independent of the arrival process).

We consider a (queueless) GI/GI/1/1-PO server
with service preemption and 
push-out (PO) when a new messge arrives.
Again, in this case $\tA\circ D = A\circ D$ so that
\be \label{DelI-def-PO}
\Delta(t) & := & A(t)-A(D(t)).
\ee
For this GI/GI/1/1-PO server, immediately after
a departure time is a server idle period. Following that will
be a busy period that concludes with a single departure.
The arrival times are renewal points, as are the (successful)
departure times.

\begin{figure}[!htb]
\centering
\includegraphics[width=2.75in]{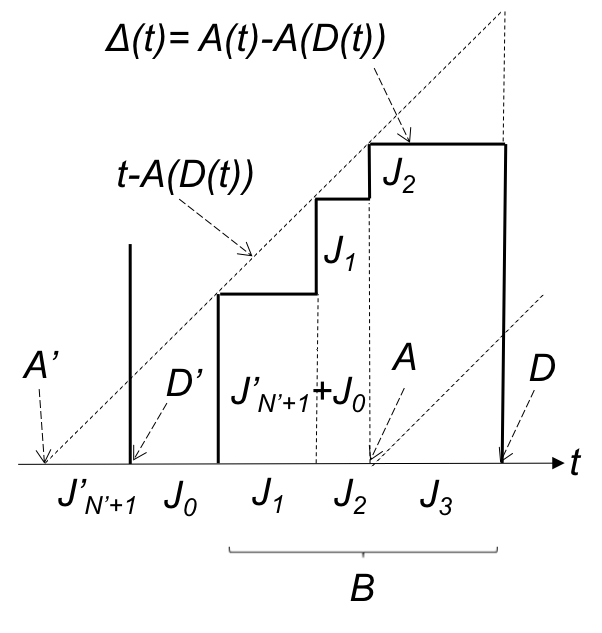}
\caption{Illustrative busy cycle $(D',D)$ for bufferless
server under push-out showing age of information
(\ref{DelI-def})
for $N=2$
(3 arrivals), server idle period $J_0$, 
interarrival times $J_1$,$J_2$, and consecutive departure
times $D',D$ (and their corresponding arrival times $A',A$),
and busy period $B=J_1+J+2+J_2$.}\label{fig:AoI-busycycle}
\end{figure}

In steady-state, 
let $B$ be distributed as the length of a busy period and
let $N$ be the number of arrivals during the busy period
not including the first arrival that starts it.
An illustrative busy cycle is depicted in 
Fig. \ref{fig:AoI-busycycle} 
for $N=2$ (3 arrivals total), where
$J_0$ is the idle period, busy period $B=J_1+J_2+J_3$,
interarrival times $J_1,J_2$, and 
completed service time $J_3$.
More generally,  the completed service time is $J_{N+1}$  and
$$B=\sum_{n=1}^{N+1}J_n.$$

Assume random variables $X,S$ are independent for the following.
The intervals
\begin{itemize}
\item $J_0\sim X-S$ given $X>S$, \ie 
$\E J_0 = \E(X-S|X>S)$
\item  $\forall n\in\{1,...,N\}$, $J_n\sim X$ given $S>X$, 
\ie $\E J_n = \E(X|S>X)$
\item $J_{N+1}\sim S$ given $X>S$, 
\ie $\E J_{N+1} = \E(S|X>S)$.
\end{itemize}

First note that $N$ is geometric with parameter 
$$\zeta := \P(X>S),$$
\ie  $\P(N=n) = (1-\zeta)^n \zeta$ for integers $n\geq 0$ and
\beqa
\E(B| N=n) & = &  n \E(X|S>X) + \E(S|X>S).
\eeqa
So, the average busy cycle duration is
\beqa
\lefteqn{ \E(D-D') = \E(J_0+B) } & & \\
& = & \E(X-S|X>S)\\
& & + 
\E\left(\sum_{n=0}^\infty (1-\zeta)^n \zeta \E(B | N=n)\right)\\
& = &  \E(X-S|X>S)\\
&  & ~+
\E N \E(X|S>X) + \E(S|X>S)\\
& = &  \E(X|X>S)
+ \frac{1-\zeta}{\zeta} \E(X|S>X)\\
& = & \E(X|X>S)+ \frac{\P(S>X)}{\P(X>S)} \E(X|S>X) \\
& = & \frac{\E X}{\P(X>S)}
\eeqa
Note that the rate at which messages are successfully transmitted
is $1/\E (D-D')$.

\subsection{D/M/1/1-PO example}

So, for the case of exponential (memoryless) $S\sim\exp(\delta)$ and deterministic
$X=1/\beta$  a.s. 
(\ie D/M/1/1-PO server), the mean rate at which messages
are successfuly transmitted is
$$\frac{1}{\E (D-D')} = \beta(1-\me^{-\delta/\beta}) =
\delta(1 - \frac{1}{2\rho} + \mbox{o}(\frac{1}{\rho})),$$
where $\rho = \beta/\delta$.

Other quantities involved can be computed as, \eg
\be
\E(S|X>S) & = & 
\int_0^{1/\beta} s\frac{\delta\me^{-\delta s}}{1-\me^{-\delta/\beta}}\dd s \nonumber \\
& = & \frac{1}{\delta}\left(1-\frac{1}{\rho}\cdot
\frac{\me^{-1/\rho}}{1-\me^{-1/\rho}}\right) 
\label{SgXgS-DM}
\ee

\subsection{M/M/1/1-PO example}

In the M/M/1/1-PO case  where
$S \sim \exp(\delta)$ and $X \sim \exp(\beta)$, we have
\beqa
\zeta & := & \P(X>S)= \frac{\delta}{\beta + \delta},
\eeqa
Also,
\beqa
\E(S|X>S) 
& = & \int_0^\infty \int_s^\infty s
\frac{1}{\zeta} 
\beta \mbox{e}^{-\beta x} 
\delta \me^{-\delta s} 
\dd x \dd s\\
& = & \frac{1}{\beta+\delta}
\eeqa
Similarly,
\beqa
\E(X|S>X) 
& =&\frac{1}{\beta+\delta}\\
\E(X|X>S) 
& = & \frac{1}{\beta} + \frac{1}{\beta+\delta}\\
\Rightarrow  
\E(X-S|X>S)   
& = & \frac{1}{\beta}
\eeqa
So, by substitution, 
\beqa
\frac{1}{\E(D-D')}
& = & 
\left(\frac{1}{\beta}+\frac{1}{\delta} \right)^{-1}
\eeqa
as expected.

\section{Stationary average $\Delta$ (\ref{DelI-def}) or
(\ref{DelI-def-PO})
for GI/GI/1/1-PO system}\label{sec:AoI}

In the following, we consider a bufferless server 
with preemption and push-out of the in-service message
when a new message arrives.

~\\
\begin{proposition}\label{prop:AoI} 
The stationary average relative age of information
of the GI/GI/1/1-PO server is
\beqa
\overline{\Delta} & = & 
\frac{\E (X \wedge S)}{\P(X>S)}
\eeqa
\end{proposition}
~\\

\noindent {\bf Proof:}
Using the Palm inversion formula \cite{Bremaud91}, 
\beqa
\overline{\Delta} & = & \frac{\E \int_{D'}^{D=D'+J_0+B} \Delta(t)\dd t}{\E(D-D')}.
\eeqa
Recall that  $\E(D-D') = \E X/\P(X>S)$.
Let $I_0=J'_{N'+1}+J_0 \sim X$ given $X>S$.
From Fig. \ref{fig:AoI-busycycle}, we see that 
\beqa
\lefteqn{\E \left(\int_{D'}^{D=D'+J_0+B} \Delta(t)\dd t 
\middle| N=n\right)} && \\
& = & \E I_0 \E B +\sum_{k=1}^n \E J_k \sum_{i=k+1}^{n+1} \E J_i  \\
& = & \E(X|X>S) (n\E(X|S>X)+ \E(S|X>S))\\
& & ~
 + \sum_{k=1}^n \Bigg((n-k)(\E(X|S>X))^2 \\
& & ~~~ + \E(X|S>X)\E(S|X>S)\Bigg)\\
& = & \E(X|X>S)\E(S|S>X)\\
& & ~+ \E(X|S>X)\E(X+S|X>S) n \\
& & ~+(\E(X|S>X))^2  \frac{n(n-1)}{2}
\eeqa
Substitute $\E N=(1-\zeta)/\zeta$, $\E N^2 = (1-\zeta)(2-\zeta)/\zeta^2$
and $\zeta:=\P(X>S)$, and simplifying the Palm inversion formula gives 
\beqa
\lefteqn{\overline{\Delta} ~=~\frac{\P(X>S)}{\E X}
\Bigg( \E(X|X>S)\E(S|X>S)  } & & \\
 & + & \E(X|S>X)\E(X+S|X>S)\frac{\P(S>X)}{\P(X>S)} \\
 & + &  (\E(X|S>X))^2  \left(\frac{\P(S>X)}{\P(X>S)}\right)^2 \Bigg)
\eeqa
Let $\tE(A|B) = \E(A|B)\P(B)$. Factoring 
$1/\P(X>S)^2$ gives
\beqa
 \overline{\Delta} 
& = & \frac{1}{\E X\P(X>S)} \Bigg( \tE(X|X>S)\tE(S|X>S)   \\
 &  & ~~+ \tE(X|S>X)\tE(X+S|X>S)  \\
 &  & ~~+ (\tE(X|S>X))^2\Bigg)  \\
 & = & \frac{1}{\E X\P(X>S)} (\tE(X|X>S)+\tE(X|S>X))\\
& & ~~	\times (\tE(S|X>S)+ \tE(X|S>X))\\
 & = & \frac{1}{\E X\P(X>S)} \E X \E (S\wedge X)
\eeqa
\qed
~\\

Now consider the definition of age of information
in Equ. (\ref{AoI-def-old}) of \cite{Yates12,Altman18}.

\begin{corollary}\label{cor:AoI-old}
For the GI/GI/1/1-PO server,
the stationary average AoI
 (\ref{AoI-def-old}) is
\beqa
\overline{{\sf AoI}} & = &\overline{\Delta}+ \frac{\E X^2}{2 \E X}.
\eeqa
\end{corollary}

\noindent {\bf Proof:}
Note that ${\sf AoI}(t) = \Delta(t) + t-A(t)$, 
where  simply by the Palm inversion formula,
the stationary average age of the latest arrival 
$\E (t-A(t)) = \E \int_0^X t \dd t/\E X$.
\qed

\subsection{D/M/1/1-PO and M/M/1/1-PO special cases for $\Delta$}

For the special case that $X=1/\beta$ a.s. and $S\sim\exp(\delta)$,
$\E(X|X>S)   =   \E(X|S>X) = 1/\beta $,
and recall (\ref{SgXgS-DM}).
So we have the following corollary of Prop. \ref{prop:AoI}.

~\\
\begin{corollary}\label{cor:DGI11-PO}
$\overline{\Delta}$  for the D/GI/1/1-PO server is
\be\label{DGI11-PO}
\E(S|\frac{1}{\beta}>S)+\frac{1}{\beta}\cdot\frac{\P(S>\frac{1}{\beta})}{\P(\frac{1}{\beta}>S)}.
\ee
In particular, 
$\overline{\Delta}$
of a D/M/1/1-PO   server is $1/\delta$.
\end{corollary}
~\\

The following corolloary states that $\overline{\Delta}$
is the same for the D/M/1/1-PO and M/M/1/1-PO special cases.

~\\
\begin{corollary}\label{cor:AoI-mm1} 
$\overline{\Delta}$  for the M/M/1/1-PO server $1/\delta$.
\end{corollary}


\bibliographystyle{plain}
\bibliography{../latex/AoI,../latex/ddos,../latex/routing}

\end{document}